# An Integrated Simulation System for Human Factors Study

Ying Wang, Wei Zhang
Department of Industrial Engineering, Tsinghua University, Beijing 100084, China

Fouad Bennis, Damien Chablat
IRCCyN, Ecole Centrale de Nantes, F44321 Nantes Cedex 3, France

## Abstract

It has been reported that virtual reality can be a useful tool for ergonomics study. The proposed integrated simulation system aims at measuring operator's performance in an interactive way for 2D control panel design. By incorporating some sophisticated virtual reality hardware/software, the system allows natural human-system and/or human-human interaction in a simulated virtual environment; enables dynamic objective measurement of human performance; and evaluates the quality of the system design in human factors perspective based on the measurement. It can also be for operation training for some 2D control panels.

## 1. Introduction

Human factors engineering requires that the design of a system should consider the human's aspects. In other word, any designed system should be user-friendly. However, in many cases, it is not easy to satisfy this requirement because it is difficult to quantify the human's requirement or the human's performance. Sometimes, iterative design at the cost of time and money is necessary. In order to reduce the iterations for time and cost concern, designers may use one of the following three approaches, or a combination of them.

The first approach is that the designers have the required quantitative data or enough clear understanding about the users. The data may include the anthropometrical data about the users, the capability of the users, and the specific scenarios of the system use. For example, by having the anthropometrical data and the scenario of use, one can design a chair that is suitable for a specific percentage users and the specific use scenario. However, in many situations, especially for new applications, having these data is not possible.

As an alternative approach, one can use physical prototypes to test the design. Now with rapid prototyping technology, one can obtain a physical prototype directly from its CAD model within one day or only hours. By testing the interaction with the user, the designer can easily find the problems and then quickly modify the design and test again. Many applications have shown that the use of rapid prototyping technology can effectively reduce the iteration time and improve the design quality [1]. However, in some cases, the designed system can be so complicated that having its prototype can be time consuming and costly. In some cases, it is even impossible.

The third approach is using a virtual prototype instead of a physical one. Arthur et al. have demonstrated that the perception, and spatial knowledge acquisition in the virtual world was indistinguishable from that of the actual physical representation of the real-world layout for certain tasks [2]. According to Buck (1998), ergonomics studies in virtual prototypes allow in early design stages to verify whether the "user interface" of a product is user friendly or not. The typical operations and procedures can be carried out in a way, which is close to reality. The design of the virtual environment and the provided virtual reality interaction tools should allow the user to make all the necessary movements. In this way, ergonomic flaws are very probable to be detected in an immersive virtual environment, and can be corrected at an early design stage, which saves a lot of effort [3]. Wilson (1999) has demonstrated that virtual reality can be a potential tool to support many types of ergonomics contribution, including assessments of office and workplace layouts, testing consequences for reach and access, reconfiguring and evaluation of alternative interface designs, checking operating or emergency procedures, and training for industrial and commercial tasks [4]. Sonoda et al. have developed a power plant training simulator based on virtual reality. The developed training simulator can realize field space as virtual space and can have a fast display of large-scale 3D models, 3D user-interfaces, and the stereoscopic vision and sound effects. The trainee can train in field operations similar to the real world [5].

The objective of our research is to develop an integrated system that incorporates virtual reality techniques for human factors study. More specifically at present, this system is for 2D control panel design evaluation and improvement, and for training. However, by having stereo projection, it is potentially possible for 3D system simulations.

## 2. System Structure

For human factors study, the developed simulation system should be able to capture the user's operation, such as the operator's hand position, gesture, and be able to recognize the specific operations (e.g. pushing a button, tuning a knob, or turning on a switch etc.). On the other hand, in order to simulate 2D control panels of different size in an actual scale, the system is required to have a screen with enough size for most applications. For accurate measurement of user's operation, the system should have quick responding features.

The main components of our hardware system are shown in Figure 1. There are four CCD cameras to detect the operator's hand position. A pair of 5DT Data Gloves is used to capture the operator's finger gesture. By combining the hand position and gesture information, the detection computer can recognize the operator's operation, such as pushing a button or tuning a knob, etc. The recognized operation is then sent to the simulation computers via LAN network. The simulation computers, depending on the screen size (width: height), may be any number. In our system, the screen size is 7.6 meters wide and 2 meters high. So we used three synchronized channels, which means there are three simulation computers and three projectors. The simulations computers work together, run the simulation scenario and the logic and displays the virtual panel to the wide screen through three 3600 lumens projectors. Then the operator can realtimely interact with the virtual control panel, such a turning on a light by pushing a certain button, or changing a quantity value by tuning a certain knob.

For simulation purpose, it is needed to have a designed control panel model. The model includes the geometric model and the control logic model. The geometric model defines the geometry of the panel as well as the textures to make it intuitive. It may be designed using any modeling software packages, such as Multigen Creator ™, 3D MAX, etc. Figure 2 shows one simplified geometric model of a control panel, with six meters, six lights, six buttons, and two display screens. The logic model defines the control logics, for example, in order to turn on a certain light, which button or buttons should be pushed. Table 1 lists part of the logics using a relation matrix. According to Table 1, when the operator pushes the "Red" button, the control logics will make "R-1" and "R-2" lights on. At the same time, he/she will hear an audio feedback, which is an alarming audio. The logics will be reflected in the developed program.

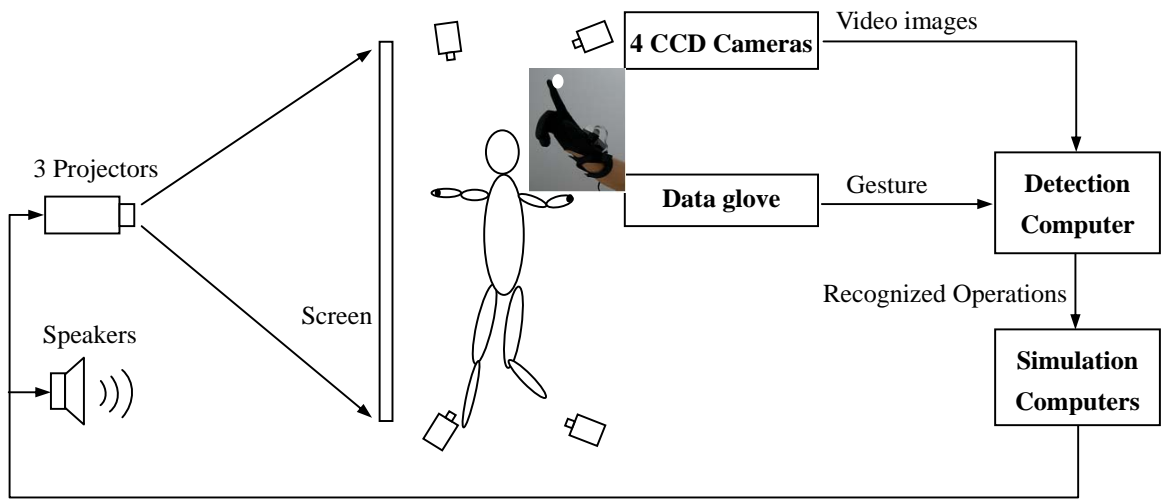

Figure 1. Hardware components of the simulation system

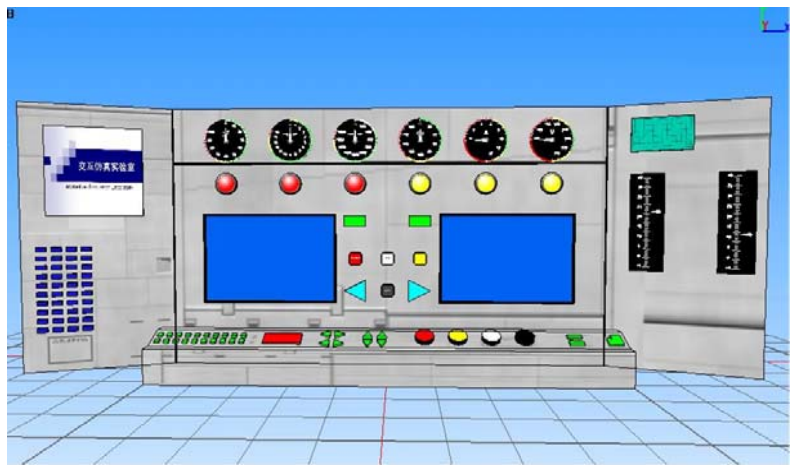

Figure 2. Geometric model of a simplified control panel

## 3. Motion Tracking and Recognition

### 3.1 Finger Position Tracking
Finger position tracking is realized by four CCD cameras. A small active light, as the position marker, is mounted on the operator's index finger. CCD apertures are reduced such that only the active lights are visible in the image, all the other objects in the environment, including the screen with projection, are not visible. This can insure enough contrast between the marker and the other objects. Before simulation, the screen area must be calibrated in order to establish the mapping relationship between the global coordinate system and the pixel coordinate system in the four CCD camera images. In other words, the marker in each specific position in the global coordinate system has a corresponding pixel coordinates in the four CCD images. In this way, during simulation, the detection computer acquires the CCD images every 40 milliseconds and computes the marker's position according to the calibrated relationship. This can be expressed as follows:

$$\begin{bmatrix} x \\ y \\ z \end{bmatrix} = \begin{bmatrix} a_{11} & a_{12} \\ a_{21} & a_{22} \\ a_{31} & a_{32} \end{bmatrix} \begin{bmatrix} i \\ j \end{bmatrix} \qquad (1)$$

where $x$, $y$, $z$ are marker's global coordinates, $a_{11}$, $a_{12}$, …, $a_{32}$ are transformation matrix which is obtained by calibration, $i$ and $j$ are marker's pixel coordinates in each CCD image [6].

Table 1. Part of the control logics

| Button | | Audio Effect | Lights | | | | | | | | Screens | | Meters |
|---|---|---|---|---|---|---|---|---|---|---|---|---|---|
| | | | R-1 | R-2 | R-3 | Y-1 | Y-2 | Y-3 | G-1 | G-2 | Left | Right | |
| Red | 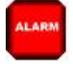 | Alarm | On | On | | | | | | | | | |
| Yellow | 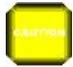 | Caution | | | | On | On | | | | | | |
| White | 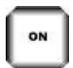 | Audio A file | On | On | | On | On | | | | | | On |
| Black | 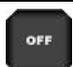 | Audio B file | Off | Off | | Off | Off | | Off | Off | Off | Off | Off |
| Left | 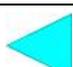 | Audio C file | | | | | | | On | | Next Slide | | |
| Right | 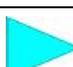 | Audio D file | | | | | | | | On | | Next Slide | |
| Default | | Env | | | Flk | | | Flk | | | | | |

* R: Red Light    Y: Yellow Lights    G: Green Lights    Env: Environment sound    Flk: Flicker

### 3.2 Finger Gesture Tracking
Finger gesture information is provided by the data glove. The glove detects each finger's gesture and expresses the gesture using a number between 0 and 255. By reading the number for each finger, the detection computer can obtain the gesture of each finger.

### 3.3 Operation Recognition
Currently, we defined three kinds of operations: pushing a button, tuning a knob, and turning on a switch. All these three operations can be recognized by analyzing the finger's position change and the gesture. They are defined as follows:

Pushing a button is TRUE when:
1) Index finger (with marker) is in a valid button zone, and
2) Index finger is open (gesture value over a certain threshold), and
3) The finger acceleration and deceleration are over a certain threshold (i.e. "pushing").

Tuning a knob is TRUE when:
1) Index finger (with marker) is in a valid knob zone, and
2) Thumb finger and the other four fingers form a "hold" gesture, and
3) The relative gesture change of the fingers forms a clockwise or counterclockwise tuning

Turning a switch is TRUE when:
1) Index finger (with marker) is in a valid switch zone, and
2) Thumb finger and the other four fingers form a "hold" gesture, and
3) The finger acceleration and deceleration are over a certain threshold (i.e. "push" or "pull").

**3.4 Operation Acquiring**

The recognized operations above are acquired in a rate of 25 times per second. The acquired data is saved both in the detection computer and in the simulation computer (server computer only, salve computers don't save). After testing, the researchers can retrieve the data and analyze. For example, how many errors did the operator make, in what kind circumstances did the operator make the errors, what is the shortest, average, and longest response time between certain events (e.g. when an alarm signal is given by a light, how long did it take for the operator to respond to the alarm).

## 4. Multi-Channel Synchronization

The simulation software is developed on the platform of Multigen Vega [TM]. Vega has a module of Distributed Vega, which enables multi-computer synchronization [7]. The synchronized computers will not compute and update the next frame if any synchronized computer has not finished the current frame. In our system, we used three simulation computers. One of the computers is defined as server, the other two as slaves. The server computer reads data transmitted from the detection computer via the network, and then distributes the data to the salves through Distributed Vega module.

Each of the three computers drives one projector for one display channel on the wide screen. Besides synchronization, it is also needed to overcome the channel overlap problem. Otherwise, the overlapped zone will have a higher brightness than the other areas. This problem is solved by channel overlap blending techniques. By applying a series of continuously changing gray bars in the overlap zone, continuous brightness is achieved along the three channels. Figure 3 shows the blended result of the simulation system. In this simulation, we used the geometric model shown in Figure 2 and the logics listed in Table 1. Due to the CCD specifications (25Hz, 768*576 pixels), currently, this system has a time resolution of 40 milliseconds (25Hz), and a spatial resolution of 0.5-2 cm on the screen, depending on the location of the screen. This performance can be further improved if high-speed CCD cameras with higher image resolutions are used. However, it is also possible to use more CCDs to increase the spatial resolutions. Time resolution seems enough for most testing applications.

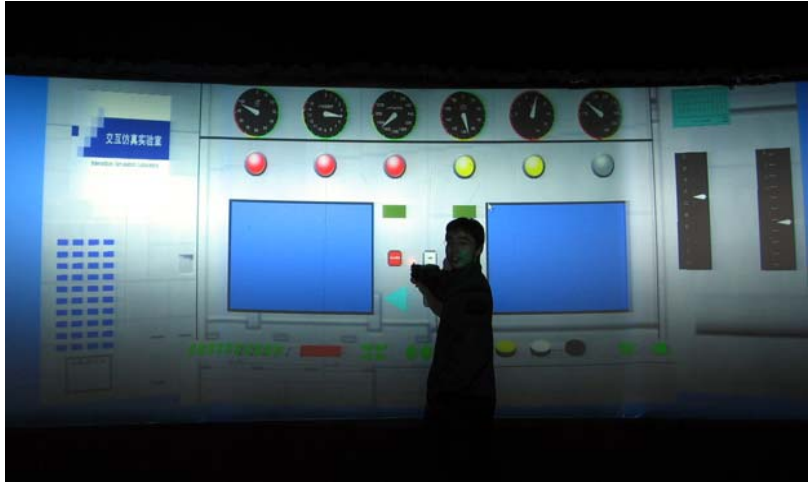

Figure 3. Multi-channel blending and simulation results

## 5. Conclusions

In this research, a wide-screen simulation system was developed. It is especially suitable for 2D control panel testing and training. Optical motion tracking and wireless data gloves eliminate the requirement of cables between the operator and computer. Distributed Vega allows synchronization of multi-computers, which enables virtually unlimited number of computers, as well as unlimited number of projectors for large screen simulation. The current stage of this research has demonstrated the feasibility of using virtual reality for human factors study. In the future, practical control panel design evaluation and training application will be tested on this system. More markers will be tracked in order to obtain more information about the whole body movement.

## Acknowledgements

The authors would like to acknowledge the financial support from the National Natural Science Foundation of China under grant number 50205014. We would also like to acknowledge the researcher exchange program in Ecole Centrale de Nantes, France, for making the collaborated study possible.

## References

[1] Dai, F. (ed.), 1998, Virtual Reality for Industrial Applications, Springer, Berlin, 2-3.
[2] Arthur, E., Hancock, P. A., Chrysler, S. T., 1997, "The Perception of Spatial Layout in Real and Virtual Worlds," Ergonomics, 40 (1), 69-77.
[3] Buck, M., 1998, "Chapter 2: Immersive User Interaction within Industrial Virtual Environment," appears in Virtual Reality for Industrial Applications, Dan, F. (ed.), Springer, Berlin, 39-60.
[4] Wilson, J.R., 1999, "Virtual Environments Applications and Applied Ergonomics," Applied Ergonomics, 30 (1), 3-9.
[5] Sonoda, Y., Yoshimoto, N., Morimura, K., Nakatani, T., Matsuzaki, S., 2000, "Development of Plant Training Simulator based on Virtual Reality," Technical Review - Mitsubishi Heavy Industry, 37(1), 19-23.
[6] Banerjee, P., Zetu, D., Virtual Manufacturing, John Wiley & Sons, 2001.
[7] http://www.multigen-paradigm.com